\documentclass{article}
\usepackage[T1]{fontenc} 
\usepackage[utf8]{inputenc} 
\usepackage{ismir,amsmath,cite,url}
\usepackage{graphicx}
\usepackage{color}
\usepackage{multirow} 
\usepackage{makecell}
\usepackage{amssymb}
\usepackage{pifont}
\usepackage{makecell}
\newcommand{\tabincell}[2]{\begin{tabular}
{@{}#1@{}}#2\end{tabular}}
\newcommand{\xmark}{\ding{55}}
\newcommand{\cmark}{\ding{51}}

\title{Efficient Adapter Tuning for Joint Singing Voice Beat and Downbeat Tracking with Self-supervised Learning Features}

\multauthor
{Jiajun Deng$^{1,2}$, Yaolong Ju$^1$, Jing Yang$^1$, Simon Lui$^1$, Xunying Liu$^2$} 
{ \\
 $^1$ Huawei Technologies Co., Ltd., Shenzhen, China\\
$^2$ The Chinese University of Hong Kong, Hong Kong SAR, China\\
{\tt\small jjdeng@se.cuhk.edu.hk, yaolongju@huawei.com}
}

\def\authorname{Jiajun Deng, Yaolong Ju, Jing Yang, Simon Lui, Xunying Liu}

\begin{document}

\maketitle
\begin{abstract}
Singing voice beat tracking is a challenging task, due to the lack of musical accompaniment that often contains robust rhythmic and harmonic patterns, something most existing beat tracking systems utilize and can be essential for estimating beats. In this paper, a novel temporal convolutional network-based beat-tracking approach featuring self-supervised learning (SSL) representations and adapter tuning is proposed to track the beat and downbeat of singing voices jointly. The SSL DistilHuBERT representations are utilized to capture the semantic information of singing voices and are further fused with the generic spectral features to facilitate beat estimation. Sources of variabilities that are particularly prominent with the non-homogeneous singing voice data are reduced by the efficient adapter tuning. Extensive experiments show that feature fusion and adapter tuning improve the performance individually, and the combination of both leads to significantly better performances than the un-adapted baseline system, with up to 31.6\% and 42.4\% absolute F1-score improvements on beat and downbeat tracking, respectively. 

\end{abstract}
\section{Introduction}\label{sec:introduction}  
Singing voice beat tracking is an important music information retrieval (MIR) task that can serve many downstream applications. For example, singing transcription can utilize beats to finetune the onsets of the transcribed notes for better accuracies \cite{nishikimi2021audio} as well as automatic accompaniment generation, where the beat information can be instrumental for drum arrangements \cite{yueh_kao_wu_2022_7316628}. However, existing literature on beat tracking mostly focused on music with instrumental accompaniment  \cite{allen1990tracking,davies2007context,mottaghi2017obtain,wu2021omnizart,di2021downbeat,goto2021musical,hung2022modeling,hernandez2022music,won2024foundation}, and tracking beats of singing voice is largely unaddressed and remains a key challenge to date. Its difficulty can be attributed to the lack of musical accompaniment that contains rhythmic and harmonic patterns vital for beat tracking in general. This leads to several challenges in developing effective singing voice beat tracking systems.  

First, the existing state-of-the-art music beat tracking systems deliver poor performances on singing voices due to the notable inherent disparities between complete music songs and singing voices \cite{heydari2022singing}. For example, the traditional music beat tracking system often learns latent mapping based on acoustic clues such as the spectrogram magnitude \cite{oliveira2010ibt,gkiokas2012music,bock2016joint}, which is often caused by the reoccurring drums or bass. Such clues, however, are barely present in singing voices. Inspired by the similarity between the singing voice and speech \cite{lindblom2014human}, the self-supervised learning (SSL) speech representations are utilized and demonstrate advantages over spectral features in singing voices \cite{heydari2022singing}. 

Second, the naturalistic singing voice data is generally highly non-homogeneous due to its inherent variabilities from different conditions, such as genres, singers, recording devices, or languages \cite{bunch1982dynamics}. The resulting high degree of singing voice heterogeneity may cause a large mismatch between training and test distributions, which can significantly degrade system performances. This issue is particularly prominent with the singing voice beat tracking that lacks musical accompaniment, as opposed to music beat tracking containing rich percussive and harmonic profiles.

To this end, we present a novel singing voice beat and downbeat tracking system using a temporal convolutional network featuring SSL representations and adapter tuning. More specifically, the SSL DistilHuBERT representations are utilized to capture the essential para-linguistics, semantic, and phonemic level characteristics and are further fused with the generic spectral features to facilitate beat estimations. A series of parameter-efficient adapters are performed to compensate for mismatch arising from the inherent variabilities among diverse singing voice datasets.
The main contributions of the paper are summarized below:

\begin{figure*}
    \centering
    \includegraphics[scale=0.7]{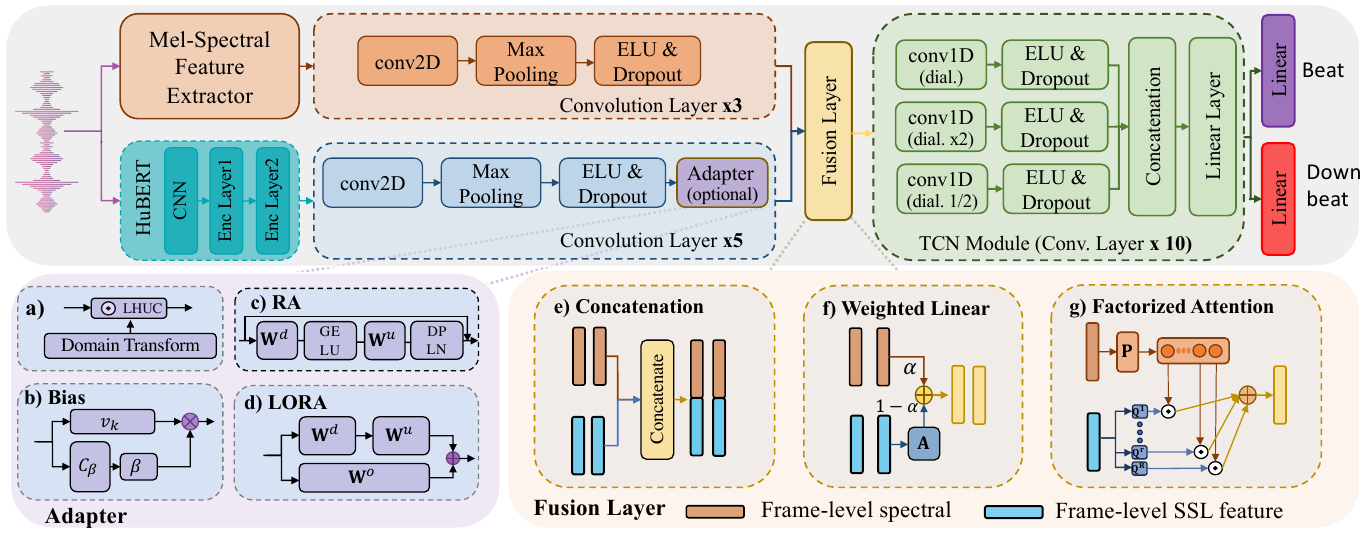}
    \vspace{-0.3cm}
    \caption{Examples of joint beat and downbeat tracking systems using Temporal Convolutional Network (TCN)  shown in the light grey box (top). The pre-trained self-supervised DistilHuBERT model is shown in the deep cyan-blue box (top left). The parameter-efficient adapters described in Section 4 are shown in the light purple colored box (bottom left corner), which includes \textbf{a)} Learning Hidden Unit Contributions adaptation, \textbf{b)} Bias Adapter adaptation, \textbf{c)} Residual Adapter (RA) adaptation and \textbf{d)} Low-Rank Adaptation (LoRA). The fusion layer described in Section 3 is shown in the light orange box (bottom right), which includes \textbf{e)} Concatenation, \textbf{f)} Weighted linear interpolation fusion, and \textbf{g)} Factorized attention fusion.}.    
    \label{fig1:tcn}
    \vspace{-0.8cm}
\end{figure*}

1) To our knowledge, this paper is the first to investigate the joint beat and downbeat tracking task featuring the fusion of SSL representations and spectral features. In contrast, similar prior research was conducted in the context of only beat estimations \cite{heydari2022singing} or beat/downbeat tracking using spectral features only \cite{heydari2023singnet}.

2) Extensive experiments show that the train-test data distribution mismatch issue presented in the non-homogeneous singing voice data significantly degrades the beat-tracking performance, particularly in downbeat estimation. To this end, inspired by the use of parameter-efficient adaptation techniques in machine learning fields \cite{sogaard2022semi,wang2022adamix,ding2023parameter,bell2020adaptation,li2023efficient,sim2024comparison}, this paper presents the first work that successfully employs efficient adapter tuning approaches for singing voice beat-tracking tasks to address the mismatch above. 

3) The efficacy of the proposed beat tracking approach is consistently demonstrated across various public datasets over the un-adapted baseline beat tracking system. In addition, the inherent generality of the proposed approach and the accompanying implementation details outlined in this paper allow their further application to other beat-tracking systems or MIR tasks.

\vspace{-0.1cm}
\section{TCN-based Beat Tracking Systems}
In this paper, we adopt temporal convolutional network (TCN) as the backbone for singing voice beat tracking for two reasons: \textbf{1)} TCN has shown solid performances in the traditional beat tracking involving musical accompaniment. First proposed by \cite{matthewdavies2019temporal}, TCN achieved superior performances to the previous SOTA bi-directional LSTM and has been widely used for beat tracking since then \cite{bock2019multi,bock2020deconstruct}. \textbf{2)} Although SpecTNT has recently outperformed TCN \cite{hung2022modeling}, TCN is still lightweight with way fewer parameters than SpecTNT, making it easy for deployment and cost-efficient as commercial applications.       

\vspace{-0.1cm}
\subsection{Architecture}
The conventional TCN-based beat tracking system consists of a front-end convolution module and a TCN module, connected by a fusion layer shown in Fig.~\ref{fig1:tcn} (light grey box, top). Each convolution layer in the front-end module has 20 channels, a stride of one, and kernels with various sizes. Max-pooling, exponential linear unit (ELU) activation \cite{singh2020elu}, and dropout neural operations are applied to each convolution layer in sequence.
The TCN module is stacked by ten dilated convolutional layers with exponentially increased dilation factors $2^0,2^1,\cdots,2^9$ resulting in a large receptive to capture long temporal contexts. Each dilated convolutional layer contains three dilated convolution blocks, each with different dilation rates (one dilation factor, half the dilation factor, and twice the dilation factor) and 20 channels. ELU activation and dropout operations are applied to each dilated convolution block, followed by an output linear layer shown in Fig.~\ref{fig1:tcn} (green box, top right). 

\subsection{Multi-task Learning}
Based on the above TCN-based architecture, the beat-tracking task can be cast as a binary classification through time, for example, classifying the presence or absence of a beat for each frame. To perform the joint beat and downbeat tracking in a single system\cite{bock2019multi,bock2020deconstruct}, an auxiliary downbeat tracking task by introducing a separate binary classification linear layer is carried out to produce the downbeat. Thus a multi-task criterion that interpolates between the beat and downbeat binary cross entropy (BCE) costs is adopted for training, which can be formulated as 
\begin{equation}
\label{eq:mtl}
{\cal L}_{MTL} =  \eta{\cal L}_{BEAT} + (1-\eta){\cal L}_{DBEAT},
\end{equation}
where $\eta\in[0,1]$ is the tunable hyper-parameter for balancing the beat BCE cost ${\cal L}_{BEAT} $ and downbeat BCE cost ${\cal L}_{DBEAT}$. Both beat and downbeat prediction outputs are post-processed with a dynamic Bayesian network (DBN) \cite{murphy2002dynamic,ellis2007beat,krebs2015efficient} to produce the final sequence of predictions \cite{bock2016joint}.

\subsection{Input Features}
Two types of feature embeddings are fed into the TCN-based beat-tracking system. The first is the traditional 81-dim log-magnitude mel-frequency spectrogram features, which are widely adopted in the music beat tracking tasks \cite{zapata2014multi,bock2020deconstruct,hung2022modeling}. Another is the SSL DistilHuBERT features. DistilHuBERT \cite{hsu2021hubert,chang2022distilhubert} is a lightweight self-supervised pre-trained speech foundation model. Its lighter architecture enables faster inference than other pre-trained foundation models. The 768-dim DistilHuBERT feature representations extracted from the last HuBERT layer are proven to serve as beneficial feature embeddings in analyzing singing rhythms \cite{heydari2022singing} due to the acoustic and linguistic similarities between singing voices and speech. 

\vspace{-0.3cm}
\section{Feature Fusion}
The process of combining diverse feature representations, named feature fusion, plays a vital role in determining the effectiveness of beat-tracking systems \cite{heydari2021beatnet}. To this end, several fusion approaches are introduced in this section to integrate the traditional spectrogram and pre-trained HuBERT feature representations effectively. 
\vspace{-0.1cm}
\subsection{Early Feature Fusion}
Early feature fusion is the combination of diverse feature representations performed early in a neural network\cite{gadzicki2020early}. For example, the features are fused at the network's input layer. Let ${{\bf x}=[\bm{x}_1,\bm{x}_2,\cdots,\bm{x}_T]}\in{\mathbb{R}^{m\times T}}$ and ${\bf u}=[\bm{u}_1,\bm{u}_2,\cdots,\bm{u}_T]\in{\mathbb{R}^{n\times T}}$ denote the spectrogram and HuBERT feature representations with $T$ frames, respectively. Two forms of early feature fusion are investigated.

\textbf{a) Input concatenation} refers to directly concatenating the spectrogram and HuBERT features at the frame level. The concatenated feature representation $\bf z$ at $t$-th frame can be expressed as ${\bm z}_{t} = [\bm{x}_t; \bm{u}_t]\in{\mathbb{R}^{n+m}}$. 

\textbf{b) Weighted linear interpolation} refers to interpolating the frame-level spectrogram and HuBERT feature representations with a learnable hyper-parameter $\alpha\in[0,1]$. The interpolated feature representation at $t$-th frame can be formulated as ${\bm z}_t = \alpha{\bm x}_t + (1-\alpha){\bf A}{\bm u}_t$,
where ${\bf A}\in{\mathbb{R}^{n\times m}}$ is a learnable projection matrix to enable the dimension of HuBERT features to be consistent with that of the spectrogram features.

\subsection{Late Feature Fusion}
The combination of diverse features at a later network layer leads to late feature fusion \cite{mungoli2023adaptive}. This allows the model to leverage high-level, abstract representations, leading to more informed decisions and improved performance. 
As shown in Fig.~\ref{fig1:tcn}, the spectrogram and pre-trained HuBERT features are first fed into a separate CNN module before being further combined using different fusion schemes. Let ${\hat{\bf x}=[\bm{\hat x}_1,\bm{\hat x}_2,\cdots,\bm{\hat x}_T]}\in{\mathbb{R}^{k\times T}}$ and ${\bf \hat u}=[\bm{\hat u}_1,\bm{\hat u}_2,\cdots,\bm{\hat u}_T]\in{\mathbb{R}^{k\times T}}$ denote the high-level CNN output hidden representations with $T$ frames using the spectrogram and HuBERT features, respectively. The concatenation and weighted combination operations can also be performed in a late fusion style, which is illustrated as \textbf{a)} late concatenation fusion ${\bm {\hat z}}_{t} = [\bm{\hat x}_t; \bm{\hat u}_t]\in{\mathbb{R}^{2k}}$ and \textbf{b)} late weighted linear interpolation fusion ${\bm {\hat z}}_t = \alpha{\bm {\hat x}}_t + (1-\alpha){\bm {\hat u}}_t$.

\subsection{Factorized Attention Fusion}
In order to focus on relevant selective representations while suppressing less important ones, factorized attention fusion is performed in a late fusion fashion. The SSL hidden representation at $t$-th frame ${\bm {\hat u}}_t$ are first factorized into $R$ subspace representations $[{\bm { v}}_t^{1}, {\bm { v}}_t^{2},\cdots,{\bm {v}}_t^{R}]\in {\mathbb{R}^{k\times R}}$ using a series of parallel linear transforms, which is expressed as 
\begin{equation}
[{\bm { v}}_t^{1}, {\bm { v}}_t^{2},\cdots,{\bm { v}}_t^{R}]=[{\bf {Q}}^{1}, {\bf {Q}}^{2},\cdots,{\bf {Q}}_t^{R}] {\bm {\hat u}}_t,
\end{equation}
where ${\bf {Q}}^r\in {\mathbb{R}^{k\times k}}$ is the linear transformations for $r$-th subspace. The spectrogram hidden embedding at $t$-th frame ${\bm{\hat x}_t}$ is projected into a $R$-dim interpolation vector $\bm{w}_t=[w_t^{1},w_t^{2},\cdots,w_t^R]\in {\mathbb{R}^{R}}$ using a projection matrix ${\bf P}\in {\mathbb{R}^{R\times k}}$, which is given as $\bm{w}_t = \text{Softmax}({\bf P}{\bm{\hat x}_t})$.
Subsequently, the fused feature representation can be obtained by an attention mechanism\cite{dai2021attentional},
\begin{equation}
{\bm {\hat z}}_t= \text{Sigmoid}(\sum_{r=1}^{R} {w_t^r}{\bm v_t^{r}}).
\end{equation}
\vspace{-0.7cm}
\section{Parameter Efficient Adaptation}
A straightforward solution to reduce the mismatch between training and evaluation distributions is to directly fine-tune the entire system using the target-domain singing voice data. However, this adaptation scheme not only encounters overfitting problems due to the scarcity of singing voices but also poses key challenges to adaptation parameter storage. Parameter-efficient adaptation approaches \cite{houlsby2019parameter,he2021towards,liu2022few} that introduce limited adaptation parameters with the original model parameters unchanged have been proposed to tail for the above overfitting and parameter storage issues. Inspired by this idea, several prominent parameter-efficient adapters are explored for singing voice beat-tracking systems in this section. 
\vspace{-0.1cm}
\subsection{Learning Hidden Unit Contributions}
Learning hidden unit contributions (LHUC) adaptation is an effective speaker adaptation solution that accounts for speaker variation of speech \cite{swietojanski2016learning}. The basic idea of LHUC adaptation is to modify the amplitudes of activation outputs using a scaling vector. Let ${\bm r}_{l,e}\in \mathbb{R}^{u}$ denote the adaptation parameters for the $e$-th domain in the $l$-th hidden layer, where $u$ is the dimension of adaptation parameters. The adapted hidden output ${\bm h}_{l,k}$ can be given as 
\begin{equation}
{\bm h}_{l,k} = {\bm h}_{l} \odot \xi({\bm r}_{l,k}),
\end{equation}
where ${\bm h}_{l}$ is the original hidden activation output at the $l$-th hidden layer, $\odot$ is the Hadamard product operation, $\xi({\bm r}_{l,e})$ is the scaling vector parameterized by ${\bm r}_{l,e}$, and $\xi(\cdot)$ is the element-wise 2$\times$Sigmoid$(\cdot)$ function with a range of $(0,2)$. During adaptation, the adaptation parameters ${\bm r}_{l,k}$ for each domain are initialized as zeros vector. An example of LHUC adaptation is shown in Fig.~\ref{fig1:tcn}(a). 

\subsection{Bias Adaptation}
The bias adapter adaptation \cite{fu2022adapterbias} adds frame-level bias to the hidden representation shifts using a domain-dependent shift vector ${\bm v}_e\in \mathbb{R}^{u}$ and a linear layer $C_{\beta}$, which is shown in Fig.~\ref{fig1:tcn}(b). The frames crucial for beat tracking should be assigned a larger representation shift compared to other frames. The linear layer produces a frame-level weight vector ${\bm \beta}=C_{\beta}{\bm h}_{l}=[\beta_1, \beta_2, \cdots, \beta_T]\in \mathbb{R}^T$, where $\beta_t$ denotes the weight of the $t$-th frame hidden representation. Therefore, the domain-dependent representation shifts ${\bm v}_e$ can be enhanced by applying frame-level weights, and the adapted hidden layer output can be expressed as 
\begin{align}
{\bm h}_{l,e} &= {\bm h}_{l} + {\bm v}_e\otimes\beta,
\end{align}
where $\otimes$ is the outer product operation, and the outer product of shift-vector $\bm v$ and the weight $\beta$ can be expressed as ${\bm v}_e\otimes\beta = [\beta_1 {\bm v}_e, \beta_2 {\bm v}_e, \cdots, \beta_T {\bm v}_e] \in \mathbb{R}^{u\times T}$.
\subsection{Residual Adapter}
Inspired by the residual idea \cite{tomanek2021residual}, a residual adapter (RA) with slight modifications is designed for beat tracking. The adapter starts with a down-linear projection ${\bf W}^d_{e} \in \mathbb{R}^{r \times u}$, followed by a non-linear GeLU activation function $\zeta(\cdot)$, and an up-linear projection ${\bf W}^u_e\in \mathbb{R}^{u \times r}$. Let $f_{RA}(\cdot;{\bm \Theta}_{l,e})$ denote the residual adapter function for $e$-th domain in the $l$-th hidden layer, where ${\bm \Theta}_{l,e}$ is the adaptation parameters for $e$-th domain. The adapted hidden outputs are given as 
\begin{align}
{\bm h}_{l,k} &= {\bm h}_{l} + f_{RA}({\bm h}_l;{\bm \Theta}_{l,e}),\\
f_{RA}(\bm{h}_{l};{\bm \Theta}_{l,e}) & = \text{LN}(\text{DP}({\bf W}^{u}_{l,e}{\zeta({{\bf W}^{d}_{l,e} \bm{h}_{l}})})),
\end{align}
where $\text{DP}(\cdot)$ and $\text{LN}(\cdot)$ denote the dropout and layernorm operations, respectively. The adaptation capacity can be controlled by managing the number of parameters in each adapter module through controlling the bottleneck dimension $r$. An example of an RA adapter is shown in Fig.~\ref{fig1:tcn}(c). 
\subsection{Low-rank Adaptation}
Instead of the non-parallel nature of adapter modules that consumes additional GPU time mentioned above, Low-rank adaptation (LoRA) \cite{hu2021lora} reduces the number of adaptation parameters by learning rank-decomposition matrix pairs $\{{\bf W}^{d}, {{\bf W}^{u}}\}$ while freezing the original weights. The LoRA-adapted linear hidden output can be expressed as 
\begin{align}
{\bm h}_{l,k} &= f_{LoRA}({\bm h}_{l-1};{\bm \Theta}_{l,e}), \\
&=({\bf W}_{l}^{o} + {\bf W}^u_{l,e}{{\bf W}^d_{l,e}}){\bm h}_{l-1},
\end{align}
where $f_{LoRA}(\cdot;{\bm \Theta}_{l,e})$ is the LoRA adapter, ${\bf W}_{l}^{o} \in \mathbb{R}^{n\times u}$ is the original pre-trained weight matrix, ${\bf W}^d_{l,e} \in \mathbb{R}^{r\times u}$ and ${\bf W}^u_{l,e} \in \mathbb{R}^{n \times r}$ are the trainable low-rank decomposition matrices. It is noted that the rank $r\ll\text{min}(u,n)$ is far less than the dimension of the original matrix, which allows for reducing the number of adaptation parameters. An example of a LoRA adapter is shown in Fig.~\ref{fig1:tcn}(d). 
\subsection{Estimation of Adaptation Parameters}
Let $\bm{\cal D}_{e}=\{{\bm X}_{e}, {\bm Y}_e\}$ denote the adaptation data for $e$-th domain, where ${\bm X}_{e}$ and ${\bm Y}_e$ stand for the singing voice waveform and the corresponding beat/downbeat sequences, respectively. Without loss of generality and for simplicity, let $\bm \Theta$ denote the original model parameters. In the context of adaptation, the adaptation parameters ${\bm \Theta}_e$ conditioned on the $e$-th domain can be estimated by minimizing the loss function in Eqn.~(1), which is given by  
\begin{equation}
\hat{\bm \Theta}_e=\arg\min_{{\bm \Theta}_e}\{{\cal L}_{MTL}(\bm{\cal D}_{e};\bm \Theta,{\bm \Theta}_e)\}.
\end{equation}
\section{Experiments}
\vspace{-0.1cm}
\subsection{Datasets and Evaluation Metrics}\label{sec:datasets}
\vspace{-0.1cm}
To the best of our knowledge, there are no publicly available datasets that include pristine vocal audio alongside beats and downbeats annotations. Annotating beats and downbeats based solely on vocal signals can be arduous and subjective, even by human experts, since there are no evident rhythmic cues like percussive instruments to accurately comprehend the singer's rhythmic intentions.

Therefore, we follow the strategy described in \cite{heydari2022singing} to utilize the existing public MIR datasets and systems to create the singing voice data with beat/downbeat annotations. This  includes \textbf{a) four music beat tracking datasets} with available beat annotations, where the singing signals are extracted by the Demucs source separation model \cite{defossez2021hybrid}, and \textbf{b) two music source separation datasets} with available isolated singing tracks, where the preliminary beats and downbeats annotations are generated by the existing TCN-based beat tracking system \cite{matthewdavies2019temporal} using the full music mix (singing with musical accompaniment), then manual revision is further performed to correct the potential  annotation errors. Altogether, six datasets are used in this paper as shown in Table~\ref{tab1:datasets}, where a silence-stripping technique is applied to each dataset to remove the long chunks of silence. The $90\%$ of the whole data randomly selected from a uniform distribution is used for training, while the remaining $10\%$ is used for evaluation.

The evaluation metric of F1-score with a tolerance window of $\pm 70$~ms, a typical setting commonly used in the traditional beat tracking \cite{matthewdavies2019temporal}, is adopted for our performance evaluation. We also adopt P-score, Cemgil, and Goto  \cite{davies2009evaluation} as additional evaluation metrics to further demonstrate the advantages of the proposed approaches in the final experiments (Table~\ref{tab3: All}). 

\begin{table}[!t]
\centering
\caption{Description of the singing voice beat tracking datasets. $\dagger$ and $*$ represent the music beat tracking dataset and the music source separation dataset, respectively.}
\vspace{0.05cm}
\label{tab1:datasets}
\resizebox{1.0\columnwidth}{!}{
\begin{tabular}{lccc} %
    \hline \hline
    \multirow{1}{*}{Dataset}  &
    \multirow{1}{*}{\# Hours}   & 
    \multirow{1}{*}{\# Excerpts}  &
    \multirow{1}{*}{Genres}  \\ \hline
    GTZAN$^{\dagger}$\cite{tzanetakis2002musical} & $5.9$ & $754$ &  Blues, Country, Disco, Hiphop, etc.  \\
    RWC Pop$^{\dagger}$\cite{de2011corpus} & $5.4$ & $273$ &  Japanese Pop., etc.  \\
    Ballroom$^{\dagger}$ \cite{gouyon2006experimental} & $2.8$ &$313$ &  Rumba, Tango, Waltz, Jive, etc.  \\
    Hainsworth$^{\dagger}$\cite{hainsworth2004particle} & $1.9$ & $173$ &  Jazz, Metal, Rock, Opera, etc. \\ 
    MUSDB18$^{*}$\cite{rafii2017musdb18} & $6.4$ & $144$ &  Pop., Country, Rock, etc.  \\
    URSing$^{*}$\cite{li2021audiovisual} & $3.4$ & $142$ &  Chinese Pop., etc.  \\\hline\hline
\end{tabular} }
\vspace{-0.4cm}
\end{table}

\vspace{-0.2cm}
\subsection{Implementation Details}
\vspace{-0.1cm}
Two feature extractors, including the mel-spectrogram feature extractor and the pre-trained SSL DistilHuBERT feature extractor\cite{chang2022distilhubert}, are employed to generate the 81-dimensional spectral features and 768-dimensional SSL feature representations of vocal signals. In this paper, the vocal signals of all datasets are resampled to 16000~Hz. As illustrated in Section 2, the temporal convolution network consists of a front-end convolution and TCN modules. The front-end convolution module tailored for late feature fusion consists of a 3-layer convolution network$\!\footnote{The channel, kernel size, stride, and padding of Conv2D used in 3-layer convolution network for each convolution layer are \{20, 20, 20\}, \{3x3,1x12,1x3\}, \{1,1,1\} and \{1x0, 0x0, 1x0\} respectively.}\!$ and a 5-layer convolution network$\!\footnote{The channel, kernel size, stride, and padding of Conv2D used in 5-layer convolution network for each convolution layer are \{20, 20, 20, 20, 20\}, \{3x3,1x12,1x3,3x3,1x12,3x3\}, \{1,1,1,1,1\} and \{1x0, 0x0, 1x0,0x0,1x0\} respectively.}\!$ for processing the spectral and SSL feature representations, respectively. The kernel size and stride for the Max-pooling operation are $1\times3$ for all convolution layers. The TCN module consists of ten dilated convolution layers, wherein the dilation factors increase exponentially. 

During the TCN-based beat tracking system training, all weights of the system are randomly initialized. The Ranger optimizer \cite{wright2021ranger21} with an initial learning rate of $0.001$, the ReduceLRonPlateau scheduler with a factor of $0.9$ and patience of $5$, and a dropout rate of $0.1$ are used for training and adaptation. The training and adaptation epochs are set as $100$ and $30$, respectively. The hyper-parameter of multi-task learning in Eqn.~(\ref{eq:mtl}) is empirically set as $\eta=0.2$. Since the ratio of positive and negative examples in the beat-tracing task is often imbalanced, the weighted binary cross-entropy loss is applied, and the weights of positive examples for the beat and downbeat costs are set to be $10$ and $20$, respectively. 

\begin{table}[!t]
\centering
\caption{Beat and downbeat tracking performance of TCN systems using different feature fusion methods evaluated on the GTZAN, RWC pop (RWCPO), and MUSDB18 (MUSDB) datasets in terms of F1-score.}
\vspace{0.15cm}
\label{tab1:feature_fusion}
\resizebox{1.0\columnwidth}{!}{
\begin{tabular}{c|c|c|ccc} %
	\hline \hline
    ID & 
    Input Features &
    Fusion Methods &
    GTZAN & RWCPO & MUSDB   \\  \hline \hline 
    \multicolumn{6}{c}{Beat/Downbeat Tracking F1 Scores} \\ \hline
    1 & Spectrogram    & -  & 0.48/0.26 & 0.65/0.53 & 0.31/0.15\\ 
    2 & DistilHuBERT   & -  & 0.74/0.47  & 0.76/0.68 & 0.38/0.17\\ \hline
    3 & \multirow{5}{*}{\tabincell{c}{Spectogram \\ \& \\ DistilHuBERT} }& Input Concatenation  & 0.78/0.56 & 0.83/0.79 & 0.41/0.25 \\ 
    4 & & Input Weighted & 0.77/0.55 & 0.86/0.81 & 0.43/0.25 \\ 
    5 & & Late Concatenation & \textbf{0.81}/0.53 & 0.87/0.81 & 0.45/0.25 \\ 
    6 & & Late Weighted  & 0.79/\textbf{0.58} & 0.88/\textbf{0.84} & 0.47/\textbf{0.26} \\
    7 & & Factorized Attention  & 0.80/0.51 & \textbf{0.91}/0.82 & \textbf{0.48}/0.23 \\ 
    \hline\hline
\end{tabular} }
\vspace{-0.4cm}
\end{table}

\begin{figure}
    \centering
    \includegraphics[scale=0.45]{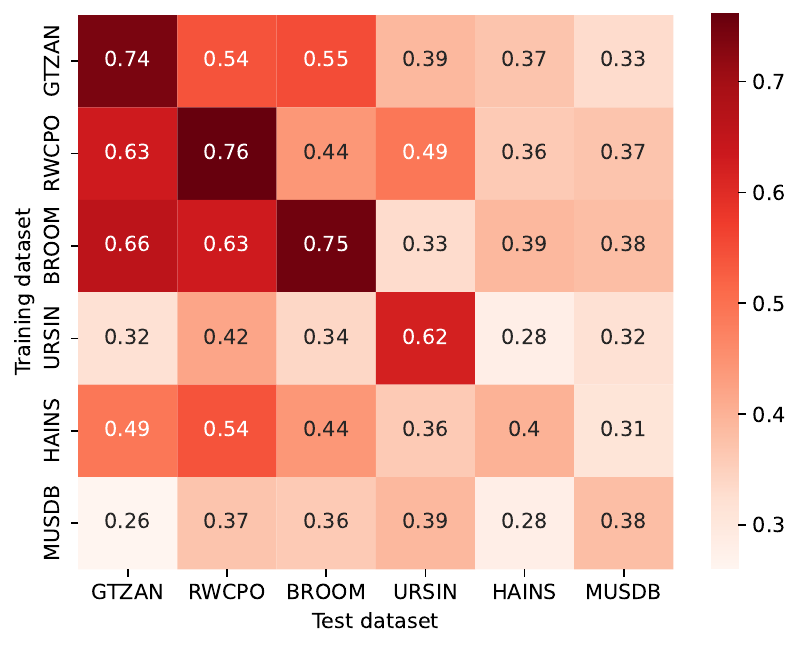}
    \includegraphics[scale=0.45]{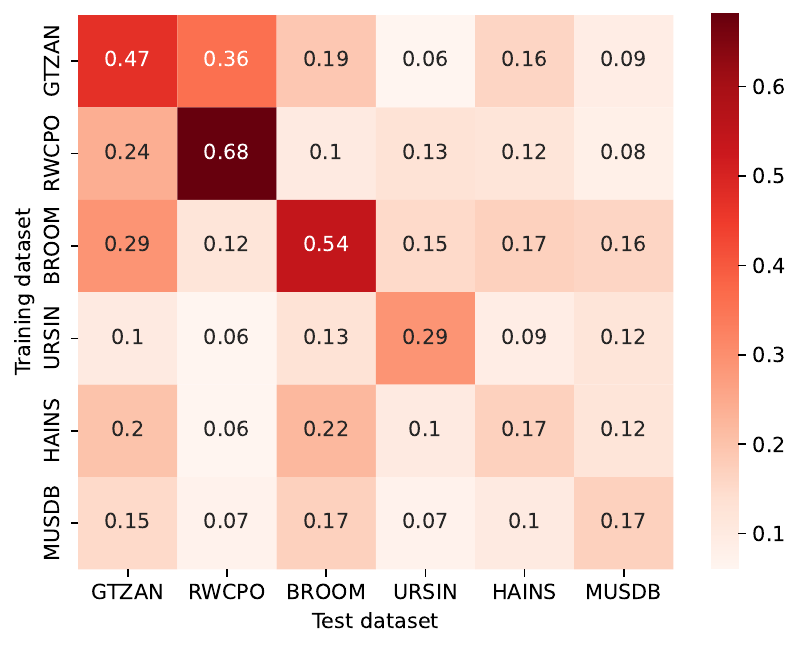}
    \vspace{-0.5cm}
    \caption{F1-score of the beat (upper) and downbeat (bottom) tracking when training data and test data are collected from the same (diagonal value) or different sources.}
    \label{fig:mismatch}
    \vspace{-0.6cm}
\end{figure}

\vspace{-0.2cm}
\subsection{Performance of Feature Fusion}
Table~\ref{tab1:feature_fusion} shows the beat and downbeat tracking F1 scores of TCN systems using different feature representation fusion methods. Several important findings can be observed. \textbf{a)} The systems using SSL DistilHuBERT features (ID.2) show better F1 performance than those using the traditional spectral features (ID.1) in all three evaluation sets. This demonstrates that the semantic information captured by SSL speech representation is crucial for singing voice beat tracking. \textbf{b)} All systems that leverage feature fusion approaches (ID.3-7) outperform the systems using only one single spectral (ID.1) or SSL feature (ID.2). This confirms our motivation that spectral and SSL features are complementary as they capture different characteristics of vocal signals. \textbf{c)} The late weighted linear interpolation fusion method (ID.6) achieves the best F1 results on the downbeat tracking and competitive beat tracking performance among all fusion approaches, therefore we selected it for the following experiments.

\begin{table*}[!t]
\centering
\caption{Beat and downbeat tracking performance of TCN systems configured with different adapters on the GTZAN, RWCPO, Ballroom (BROOM), Hainsworth (HAINS), MUSDB, and URSing (URSIN) datasets in terms of F1-score.}
\vspace{0.1cm}
\label{tab2: adapter}
\resizebox{1.98\columnwidth}{!}{
\begin{tabular}{c|c|lll|cccccc} %
    \hline \hline
    \multirow{2}{*}{ID} & 
    \multirow{2}{*}{Systems} &
    \multicolumn{3}{c|}{Adapter} &
    \multicolumn{6}{c}{Datasets}  \\   
    \cline{3-11}
    & & Method & Location & \# Params & GTZAN & RWCPO & BROOM &  HAINS & MUSDB & URSIN \\ \hline \hline
    \multicolumn{11}{c}{Beat/Downbeat Tracking F1-scores} \\ \hline\hline
    1 & In-domain  & -& - & -  &0.74/0.47 &0.76/0.68  &0.75/0.54 &0.40/0.17 &0.38/0.17 &0.62/0.29\\ 
    2 & Multi-condition & - & - & -  &0.78/0.57 &0.89/0.45  &0.72/0.43 &0.50/0.33 &0.53/0.25 &0.61/0.22 \\ \cline{1-11}
    3 & \multirow{7}{*}{\tabincell{c}{Multi-condition  \\ with \\ Adaptation} } & Fine-tune & ALL Layers & 100\%  &\textbf{0.80}/0.59 &\textbf{0.91}/\textbf{0.82}  &\textbf{0.80}/\textbf{0.57} &0.54/0.38 &0.55/0.38 &\textbf{0.72}/0.34 \\
    \cline{3-5}
    4 & & LHUC & First CNN Layer  & 5\%  &0.78/0.56 &0.88/0.59  &0.72/0.46 &0.50/0.33 &0.53/0.26 &0.62/0.25  \\
    5 & & BIAS & First CNN Layer  & 10\%  &0.79/0.57 &0.89/0.61  &0.71/0.47 &0.52/0.35 &0.55/0.31  &0.62/0.24 \\
    6 & & LoRA & First CNN Layer & 20\% &\textbf{0.80}/0.62 &0.90/0.68 &0.78/0.56 &0.54/0.37  & 0.56/0.38  &0.66/0.33 \\
    7 & & RA &  First CNN Layer & 20\%  &\textbf{0.80}/\textbf{0.65} & \textbf{0.91}/0.80   &\textbf{0.80}/\textbf{0.57} &\textbf{0.58}/\textbf{0.43} &\textbf{0.58}/\textbf{0.41} & 0.68/\textbf{0.38} \\\cline{3-5}
    8 & & RA & Second CNN Layer  & 10\% &0.78/0.60 &0.90/0.77  &0.76/0.56 &0.57/0.41 &0.55/0.40 &0.68/0.35 \\
    9 & & RA & Third CNN Layer & 5\% &0.79/0.62 &0.90/0.78 &0.76/0.56 &0.57/0.39 &0.55/0.40 &0.68/0.33  \\ \hline \hline
    

\end{tabular} }
\vspace{-0.4cm}
\end{table*}

\begin{table*}[!t]
\centering
\caption{Beat and downbeat tracking performance of the proposed TCN systems incorporating fusion and adapters.}
\vspace{0.1cm}
\label{tab3: All}
\resizebox{2\columnwidth}{!}{
\begin{tabular}{c|c|c|c|c|cccc|cccc} %
    \hline \hline
    \multirow{2}{*}{ID} & 
    \multirow{2}{*}{Systems} &
    \multirow{2}{*}{Input Features} &
    \multirow{2}{*}{\tabincell{c}{Feature \\ Fusion}} &
    \multirow{2}{*}{Adapter} &
    \multicolumn{4}{c|}{Beat Tracking} &
    \multicolumn{4}{c}{Downbeat Tracking}\\   
    \cline{6-13} 
    & & & & &  F1 & P-score & Cemgil & Goto &  F1 & P-score & Cemgil & Goto\\ \hline \hline
    1 & Multi-condition & Spectrogram & \xmark & \xmark & 0.497 &0.541 & 0.410 & 0.429 &0.254 & 0.420 &0.212 & 0.256 \\
    2 & Multi-condition & DistilHuBERT & \xmark & \xmark & 0.656 &0.681 & 0.565 & 0.561 &0.389 & 0.501 &0.351 & 0.370 \\
    3 & Proposed & Spec. \& HuBERT & \cmark & \xmark & 0.774 &0.756 & 0.684 & 0.703 &0.524 & 0.603 &0.487 & 0.520 \\
    4 & Proposed & Spec. \& HuBERT & \cmark & \cmark & \textbf{0.813} &\textbf{0.801} & \textbf{0.713} & \textbf{0.757} &\textbf{0.678} & \textbf{0.692} &\textbf{0.621} & \textbf{0.663} \\
    \hline\hline
\end{tabular} }
\end{table*}

\subsection{Performance of Adaptation}
The mismatch across different datasets is revealed in Fig.~\ref{fig:mismatch}. \textbf{a)} When the system is trained on singing voice data from the same source as the test data, the best beat and downbeat tracking performance are obtained (diagonal value). \textbf{b)} The mismatch between training and test distributions (non-diagonal) significantly degrades the beat tracking performance, especially in downbeat tracking. This confirms our assumption that the mismatch across different datasets is particularly prominent in singing voice beat tracking due to the lack of robust rhythmic and harmonic patterns. Therefore, mismatch is an essential issue that needs to be addressed in multi-condition training.

Table~\ref{tab2: adapter} shows the beat and downbeat tracking performance of the TCN systems configured with different adapters using only DistilHuBERT features. Several trends can be found. \textbf{a)} It is not surprising that multi-condition systems (ID.2) trained on all six datasets do not always outperform the in-domain systems (ID.1) trained on the well-controlled data from the same source as test data because of the mismatch issue. This demonstrates that blindly expanding the training data is insufficient to enhance the system's generalization. \textbf{b)} All systems configured with adapters (ID.4-7) improve the performance over both multi-condition systems (ID.2) and in-domain systems (ID.1), which suggests that parameter-efficient adapter tuning methods can address the mismatch issue effectively. \textbf{c)} The residual adapter (RA) (ID.7) applied at the first CNN layer achieves the best results relative to other adaptation approaches. It is noteworthy that RA adaptation using only $20\%$ of adaptation parameters shows comparable performance to fully fine-tuned techniques (ID.3). In addition, the observation that the performance gain of downbeat tracking is greater than that of beat tracking is consistent with our finding in Fig.~\ref{fig:mismatch} that the downbeat tracking performance is more sensitive to the mismatch issue. \textbf{d)} When incorporating adapters into the second or third CNN layer (ID.8,9), with acceptable performance degradation, RA adaptation delivers a much lighter architecture with fewer parameters.
\vspace{-0.3cm}
\subsection{Performance of The Proposed Method}
\vspace{-0.1cm}
The advantages of the proposed method incorporating both late weighted linear interpolation feature fusion and RA adapter are demonstrated in Table~\ref{tab3: All}. The evaluation results are the overall performance of all six evaluation sets using micro averaging. Two main observations can be found. 

First, the multi-condition system using the proposed feature fusion approaches (ID.3) still outperforms the systems (ID.1,2) using only one spectral or SSL feature. Of particular interest, this system (ID.3) is compared to the existing singing voice beat tracking system \cite{heydari2022singing}, where the same evaluation protocol is followed by using the entire $5.9$-hrs GTZAN dataset for testing$\!\footnote{The remaining five datasets are therefore used for training our system, which is less data compared to \cite{heydari2022singing}.}\!$. As a result, our system achieved beat tracking F1-score of 0.784 on GTZAN, a significant $\textbf{5.1\%}$ absolute improvement compared to 0.733 from \cite{heydari2022singing} even using less training data. 

Second, consistent performance improvements across all evaluation metrics are observed when the adapter tuning scheme (ID.4) is applied. Overall significant F1-score improvements of up to $\textbf{31.6\%}$ and $\textbf{42.4\%}$ absolute were obtained over the baseline un-adapted system using only one feature on the beat and downbeat tracking, respectively. In particular, the beat/downbeat performances of $0.87/0.78$, $0.95/0.87$, $0.85/0.75$, $0.66/0.49$, $0.68/0.49$, and $0.79/0.41$ are achieved by the proposed approach (ID.4) on the test split of GTZAN, RWC pop, Ballroom, Hainsworth, MUSDB, and URSing datasets, respectively. 
\vspace{-0.3cm}
\section{Conclusions}
This paper proposed a temporal convolution network based beat-tracking framework featuring self-supervised learning (SSL) representations and efficient adapter tuning to track the beat and downbeat of singing voices jointly. Feature fusion strategies were performed to leverage the advantages of the generic spectral and SSL speech feature representations. Efficient adapter tuning was utilized to mitigate the sources of variabilities of the non-homogeneous singing voice data. Experimental results showed that the proposed approach significantly outperforms the un-adapted baseline system using only spectral or SSL features. The inherent generality of the proposed approaches allows their further application to other beat-tracking systems or MIR tasks. Future work will focus on solving the data sparsity issue of the singing voice beat tracking task. 

\vspace{-0.3cm}
\section{Acknowledgments}
This research is supported by Hong Kong RGC GRF grant No. 14200021, 14200220, Innovation \& Technology Fund grant No. ITS/254/19 and ITS/218/21. Jiajun Deng carried out this research during his internship at Huawei Hong Kong Research Center. The authors would like to thank Huawei for its hardware and software services.





\bibliography{ISMIRtemplate}

%
%
%
%
%


\end{document}